\documentclass[prb,amsfonts,a4paper,floatfix]{revtex4-1}

\usepackage{eurosym}
\usepackage{graphicx}
\usepackage[export]{adjustbox}

\usepackage[utf8]{inputenc}

\usepackage[sort&compress]{natbib}
\bibpunct{[}{]}{,}{n}{}{,} 

\usepackage{bm}          
\usepackage{amsmath}
\usepackage{latexsym}
\usepackage{color}

\graphicspath{{./Figures/}}

\newcommand{\HH}{{\cal H}}
\newcommand{\HHo}{{\cal H}_0}

\newcommand{\vKS}{\ensuremath{v^{\mathrm{KS}}}}
\newcommand{\vHXC}{\ensuremath{v^{\mathrm{HXC}}}}

\newcommand{\braket}[1]{{\langle{#1}\rangle}}

\newcommand{\e}{{\mathrm{e}}}

\begin{document}

\title{Learning DFT}

\author{Peter Schmitteckert}
\affiliation{HQS Quantum Simulations GmbH\\ 
Haid-und-Neu-Straße 7\\
76131 Karlsruhe\\
Germany}
\date{\today}

\begin{abstract}
We present an extension of reverse engineered Kohn-Sham potentials from a density matrix renormalization group
calculation towards the construction of a density functional theory functional via deep learning. 
Instead of applying machine learning to the energy functional itself, we apply these techniques to the Kohn-Sham potentials.
To this end we develop a scheme to train a neural network to represent the mapping from local densities to Kohn-Sham potentials.
Finally, we use the neural network to up-scale the simulation to larger system sizes.
\end{abstract}

\maketitle

\section{Introduction}

In the quest of solving strongly interacting quantum systems the density matrix renormalization group technique
(DMRG)  \cite{White:PRL92,White_Noack:PRL92,White:PRB93,Noack_Manmana:AIPCP2005,Hallberg:AP2006}  has turned out to be a powerful tool.
Representing a many-particle wave function based approach it is perfectly suited to attack strongly interacting
many particle problems. However, its downside are the high computational costs, rendering the DMRG beeing too expensive
for large systems. In contrast, the density functional  theory (DFT) 
has proven to be successful in the prediction of structure  of molecules, solids and surfaces,\cite{HohenbergKohn:1964,KohnSham:1965,dreizler90}
although it is based on a single particle description only.
The DFT is based on the Hohenberg-Kohn theorem \cite{HohenbergKohn:1964} which states that the ground state properties of
a many-particle system is determined by the local density, specifically, for each observable there exists a functional of the density,
which provides the ground state expectation value of the observable, when evaluated with  the ground state density.
In result the ground state energy functional is minimized by the ground state energy.
Within the Kohn-Sham construction \cite{KohnSham:1965} one describes an interacting system by a non-interacting system,
where the local Kohn-Sham potential $\vKS_x$ replaces the local potential to mimic the effect of the interaction.
Provided the interaction is kept fix there is a one-to-one correspondence between the local potentials and the local densities.
Despite this solid foundation of the DFT and decades of research, DFT is unreliable for strongly correlated electron systems,
as the nature of the associated functional is unknown.
Gunnarson and Sch\"onhammer \cite{SchoenhammerGunnarsson:PRL1986} extended the DFT 
to a homogeneous lattice system by explicitly reverse engineering the Kohn-Sham potentials
of the solution provided by exact diagonalization . 
This approach was the extended in \cite{Schmitteckert_Evers:PRL2008} to inhomogeneous systems, see also \cite{PS:PCCP2018}.
In this work we study the extension of the reverse engineered Kohn-Sham potentials for specific systems 
to the construction of DFT functionals.

To achieve such a construction we make use of deep learning and neuronal networks (NN), where we refer to
the excellent introduction by Nielson\cite{Nielsen:DP2015} and Appendix~\ref{sec:FitNN}.
NN have actually a long history, starting with the pioneering work of McCulloch, Warren, and Pitts \cite{McCulloch:BMB1943}.
The major steps in the development of today's  success in pattern recognition consists in the development of the
back propagation algorithm \cite{Rumelhart:Nature1986} and the invention of convolutional networks \cite{LeCun:1995},
combined with the computational power provided by graphic cards.
In addition, the availability of free software packages\cite{tinyDNN,tensorflow:2015,keras:2015} simplifies the application
of neural network enormously.

Machine learning has already a broad application in physical simulations, for a review see\cite{RevModPhys.91.045002}.
In the context of simulating electronic systems machine learning has been applied to
bypass the Schr\"odinger equation\cite{Brockherde:NC2017,Kolb:SR2017}. That is, training the neural nework with DFT and post-DFT results
in order to predict properties directly. 
Another approach consists in improving existing DFT functionals for molecules \cite{Hu:JCP2003,ZHENG:CPL2004,Liu:JPCA2017}.
In addition, DFT functionals have been constructed by learning the energy functional.\cite{Snyder:PRL2012,Snyder:JCP2013,Burke:PRB2016}
Here we are following a different route. Instead of applying machine learning to the energy functional, we apply it
to the learning of the Kohn-Sham potentials. The idea of this approach is that it should simplify a divide and conquer 
approach to larger, even inhomogeneous systems. In addition, there is no problem associated with the functional derivatives 
of the energy functional, as we are already learning the derivatives.

In this work we look at a one dimensional interacting Fermi system with disorder Eq.~\eqref{eq:H}.
The model has been studied for a long time and caught attraction by the work of Giamarchi and Schultz \cite{GiamarchiSchulz:PRB1988},
who predicted a phase transition from the Anderson insulator to a metallic Luttinger liquid for  attractive interaction, $U=-1t$,
and weak disorder. For repulsive interaction the interaction enhances the localization induced by the 
disorder\cite{GiamarchiSchulz:PRB1988,PS:PRL1998} while for strong disorder and interaction the interplay
of disorder and charge density wave ordering renders the system complicated\cite{PS:PRL1998b,Jalabert:PE2001}.

In Section~\ref{sec:model} we introduce the model under investigation and discuss the application of a NN to the system in 
Section~\ref{sec:NNDFT} and the up-scaling is presented in Section~\ref{sec:FSC}.
In Appendix~\ref{sec:FitNN} we provide a detailed introduction into fitting functions with neural networks and finally
sketch an extension using convolutional networks in Appendix~\ref{sec:CNN}.

\section{The model}
\label{sec:model}
The model under investigation is chosen to be formally simple, but beyond the reach of standard DFT functionals.
To this end we look at spinless fermions in one dimension, with a nearest neighbor hopping $t=1$, periodic boundary conditions (PBC),
nearest neighbor interaction $U$ and a strong onsite disorder $v_x$,
\begin{align}
 \HH &=     \underbrace{-t \,\sum_x  \hat{c}^\dagger_{x-1} \hat{c}^{}_x  \;+\;  \hat{c}^\dagger_{x} \hat{c}^{}_{x-1} }_{\text{kinetic part}}
     \;+\;  \underbrace{U  \,\sum_x  \hat{n}^\dagger_{x-1} \hat{n}^{}_x}_{\text{interaction}}
     \;+\;  \underbrace{\sum_x v_x \hat{n_x}}_{\text{disorder}}
     \,.\label{eq:H}
\end{align}
Here, $\hat{c}^{\dagger}_x$ ($\hat{c}^{}_x$) denote the fermionic creation (annihilation) operator at site $x$,
$\hat{n}^{}_x = \hat{c}^{\dagger}_x\,\hat{c}^{\vphantom{\dagger}}_x$ the local density, and $M=98$ the number of lattice sites.
The interaction is chosen to be $U=1.0$, and the disorder potentials $v_x$
is choosen from a uniform distribution between $\pm W/2$, and smoothened with a cosine filter with a width of 3 three sites.
Specifically, we choose $\epsilon_x \in [-W/2, W/2]$ uniformly distributed and obtain the smoothened
potentials from
\begin{align}
 v_x = \sum_{s=-d}^{s=d}  \cos^2\left( \frac{ s\,\pi }{ 2 d+2} \right) \, \frac{ \epsilon_{x+s}}{ \sqrt{d+1}}
\end{align}
with $d=3$ and $W=2$. The reason for the smoothening is that disorder typically stems from scatterer in the substrate.
So each scatterer should also effect neighboring sites.
From test runs with fewer samples than provided below, but without the smoothing, we conclude
that it doesn't alter the findings of this work.

In order to obtain the reference data for training the network we perform a sample statistics
using the density matrix renormalization group technique
where we used the ``safety belt'' approach as explained in \cite{PS:Proceedings98,Proceedings98} to avoid
getting trapped in an excited state during the DMRG initialization.

\section{Neural network as a generator of DFT potentials}
\label{sec:NNDFT}

From each DMRG run we obtain the local density $n_x$.
We then perform an inverse DFT \cite{SchoenhammerGunnarsson:PRL1986,SchoenhammerGunnarssonNoack:PRB1995,Schmitteckert_Evers:PRL2008,FE_PS:PSS2013,PS:PCCP2018},
where we search for a non interacting Hamiltonian,
\begin{align}
 \HHo &=     \underbrace{-t \,\sum_x  \hat{c}^\dagger_{x-1} \hat{c}^{}_x \;+\;  \hat{c}^\dagger_{x} \hat{c}^{}_{x-1}}_{\text{kinetic part}}
     \;+\;  \underbrace{\sum_x \vKS_x \,\hat{n}_x}_{\text{Kohn-Sham potential}}
     \,,\label{eq:H0}
\end{align}
that is we search numerically for the Kohn-Sham potential $\vKS_x$, such that 
\begin{align}
    \braket{\hat{n}_x}_{\HH} &= \braket{\hat{n}_x}_{\HHo} \,.
\end{align}
In return we obtain for each disorder realization $\{v_x\}$ the corresponding Kohn-Sham potentials
$\vKS_x$. By building an infinite table listing the Kohn-Sham potential for every possible disorder configuration
we would obtain the full DFT functional for this system. Since this is not feasible we explore the possibility
of using a NN to construct such a functional. We would like to remark, that we are actually constructing the Kohn-Sham
potentials and the energy functional is given by solving the Kohn-Sham system.

At a first sight it looks rather boring to construct a DFT functional for a system that one has already solved
using the DMRG. The main reason for this work is that we would like to apply the functional constructed in this
section for larger systems, i.e. to perform an up-scaling in system size, see Sec.~\ref{sec:FSC}.

In our setup we constructed a set of training and test data by performing an ensemble statistics
for 14.950 systems for the training set, and 50 realizations for the testing set, where the testing
set was never used as a training input. All simulations are performed at half filling.
In order to avoid getting stuck at an excited state during the DMRG sweeping we added the
ground state of a homogeneous, delocalised systems to the density matrix used for the selection of the 
target space.\cite{PS:Proceedings98,PS:PRL1998,PS:PRL1998b} The reason for this approach is explained in detail in \cite{PS:Proceedings98}.
In addition we targeted for the three states lowest in energy in an initial run, performed three finite lattice sweeps
and keept at least 400 states per block, while we increased  the number of states per block to ensure a discarded entropy below $10^{-5}$.
We then restarted each run keeping at least 450 states per block, targeting for the two states lowest in energy
and performing three sweeps and ensured a discarded entropy below $10^{-6}$. 
Finally we restarted each run again, keeping at least 500 states per block, targeting
the ground state only, performed seven finite lattice sweeps and inreased the number of states kept per block
to ensure a discarded entropy below $10^{-8}$.

For each sample the calculated the corresponding Kohn-Sham potentials $\vKS_x$. 
For each site $x$ of a system  we constructed a data set consisting of  
the densities starting $s=35$ sites left to $x$ up to the densities for the $s$ sites right of $x$, employing PBC,
\begin{align}
 \{ n_{x-s}, n_{x-s+1}, \cdots , n_{x+s} \} &\longrightarrow \vKS_x   \label{eq:set}
\end{align}
mapping each set of densities to the corresponding Kohn-Sham potential $\vKS_x$.
We didn't use the complete set of $M=98$ densities as we later want to use the data for up-scaling
and we wanted to avoid targeting at precisely $M=98$ sites.
We therefore obtain $98 * 14.950 = 1.465.100$ training sets Eq.~\eqref{eq:set} and $98 * 50 = 4.900$ testing sets Eq.~\eqref{eq:set}
with an input length of $71 = 2 * 35 + 1$ and a single valued result.
We applied a $\tanh$ activation and therefore re-scaled the input densities by $ 2\left( n_x - 0.5 \right)$.
The network used has as $71 \times 251 \times 249 \times 1$ structure, that is 71 input values, namely the densities,
one output values, the Kohn-Sham potential, two hidden layers consisting of 251 and 249 neurons.
We have performed tests with other layouts, especially with deeper networks. The results obtained with
those networks were very similar to ones presented below.

For details on how to fit functions with NN we refer to appendix \ref{sec:FitNN}.

We trained the network first by using the {\tt tiny DNN} \cite{tinyDNN} software package and switched later to
{\tt tensorflow} \cite{tensorflow:2015} with the {\tt Keras} \cite{keras:2015} interface.
During the training phase we tried various available optimizer, most notably
stochastic gradient descent ({\tt SGD}) \cite{robbins:AMS1951}, 
adaptive moment estimation ({\tt Adam}) \cite{kingma2014adam}, 
and the adaptive gradient algorithm {\tt Adagrad} \cite{Duchi:JMLR2011}. 
In this section we are only reporting the results
for the optimization run with the lowest deviation which we achieved using {\tt tiny DNN}.

In addition to the Kohn-Sham potentials $\vKS_x$ we are also defining the
Hartree exchange correlations potentials $\vHXC_x$
\begin{align}
  \vHXC_x = \vKS_x - v_x \,
\end{align}
that is, the interaction induced change of the local potential. Note that in the non-interacting limit, $U=0$,
the Hartree exchange correlation potential is zero by definition. We did not subtract the Hartree contribution
from $\vHXC$ as the Hartree-Fock approximation has problems for strongly correlated one-dimensional systems, see \cite{PS:PCCP2018}.

\begin{figure}[ht]
\begin{center}
(a)\;\includegraphics[width=0.4\textwidth,valign=t]{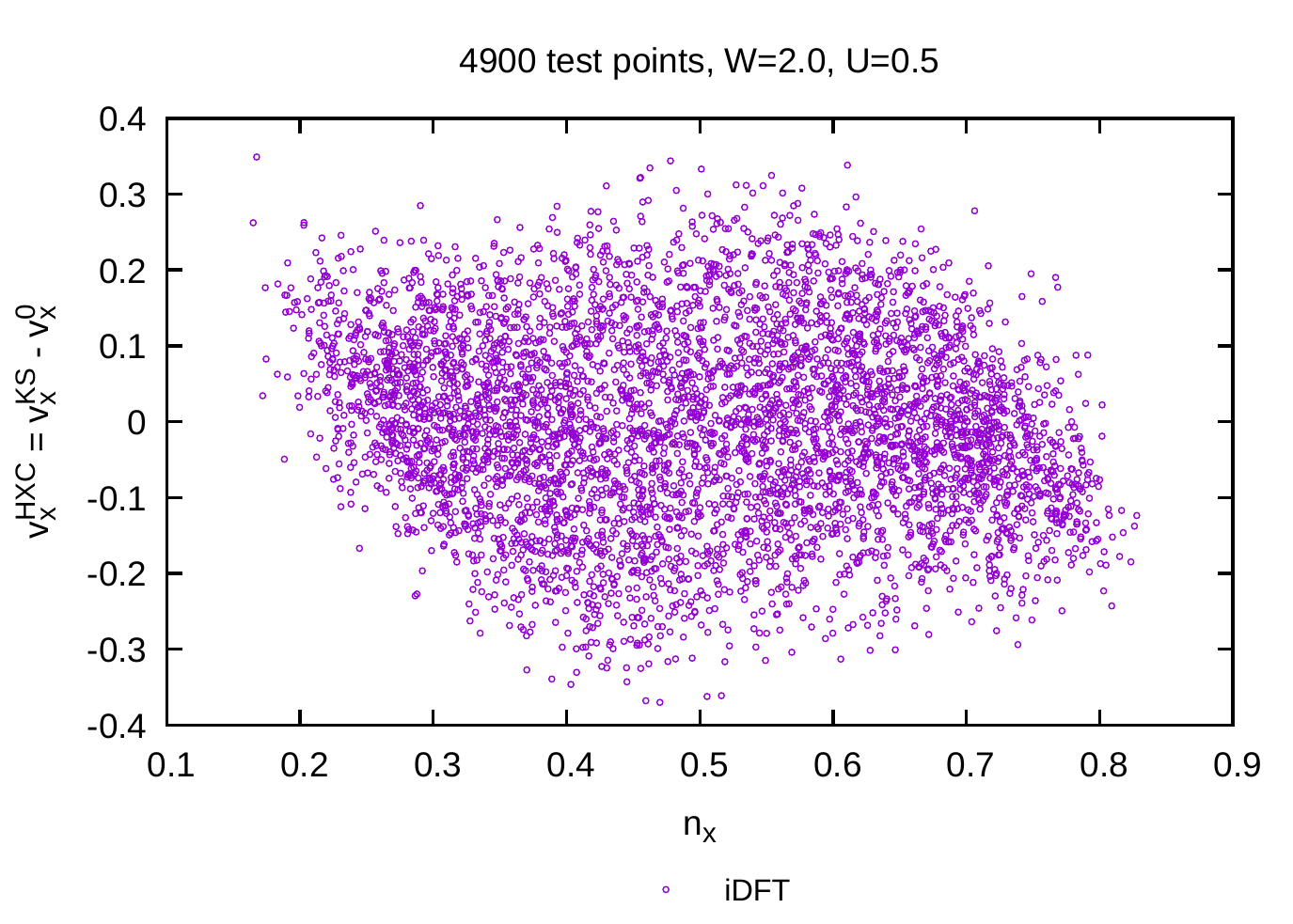} 
\qquad
(b)\; \includegraphics[width=0.4\textwidth,valign=t]{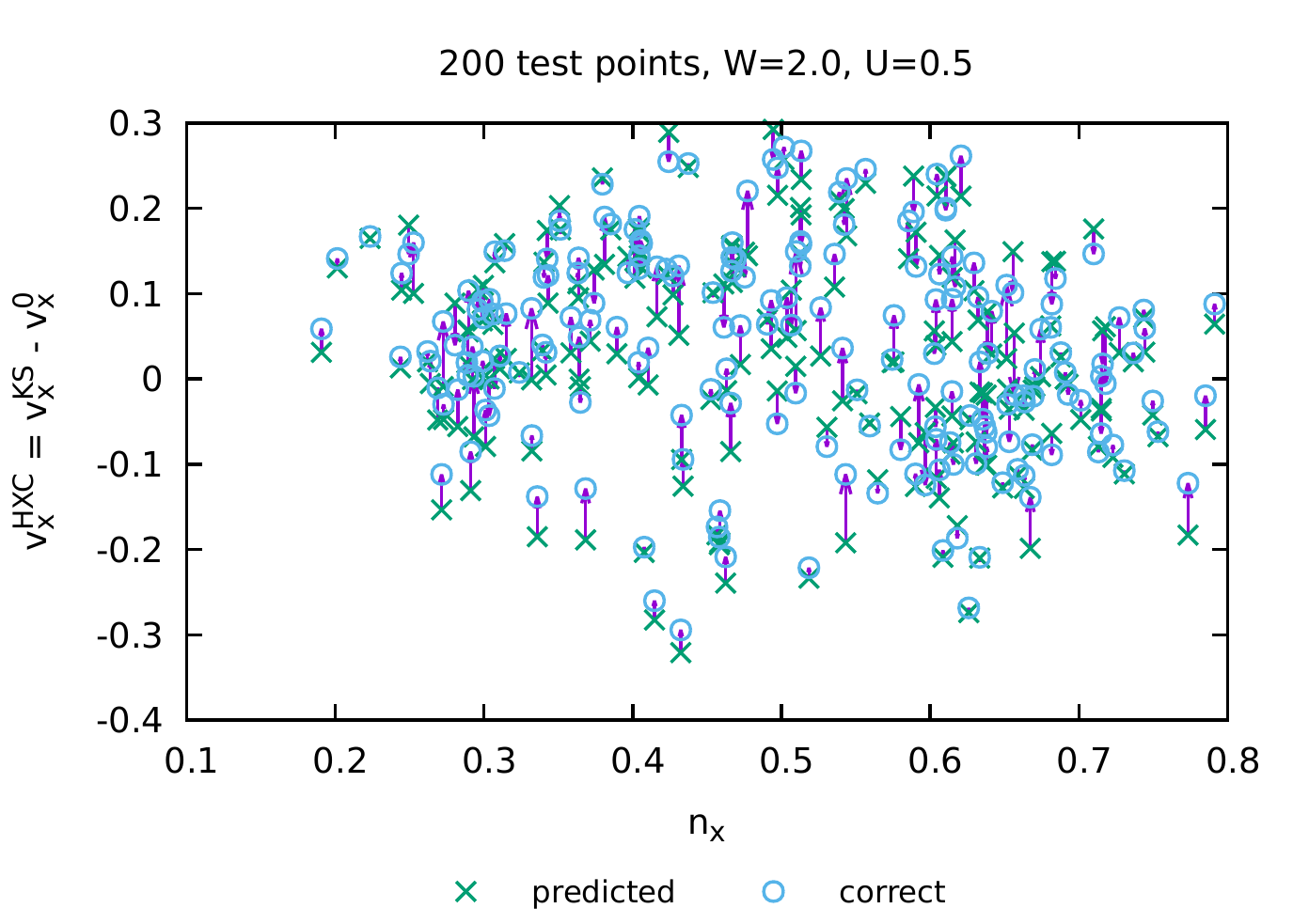}
\\
(c)\;\includegraphics[width=0.4\textwidth,valign=t]{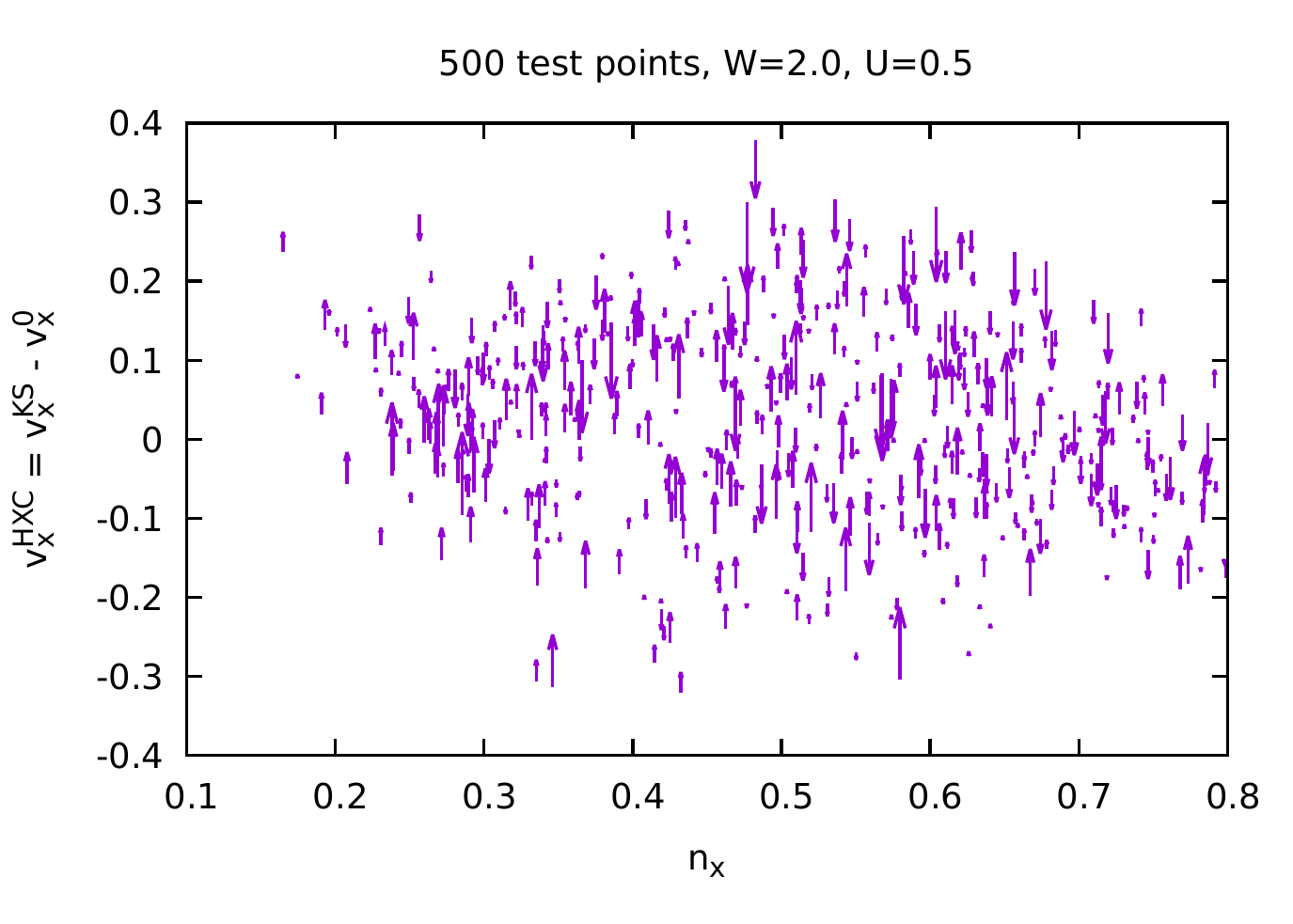}
\qquad
(d)\; \includegraphics[width=0.4\textwidth,valign=t]{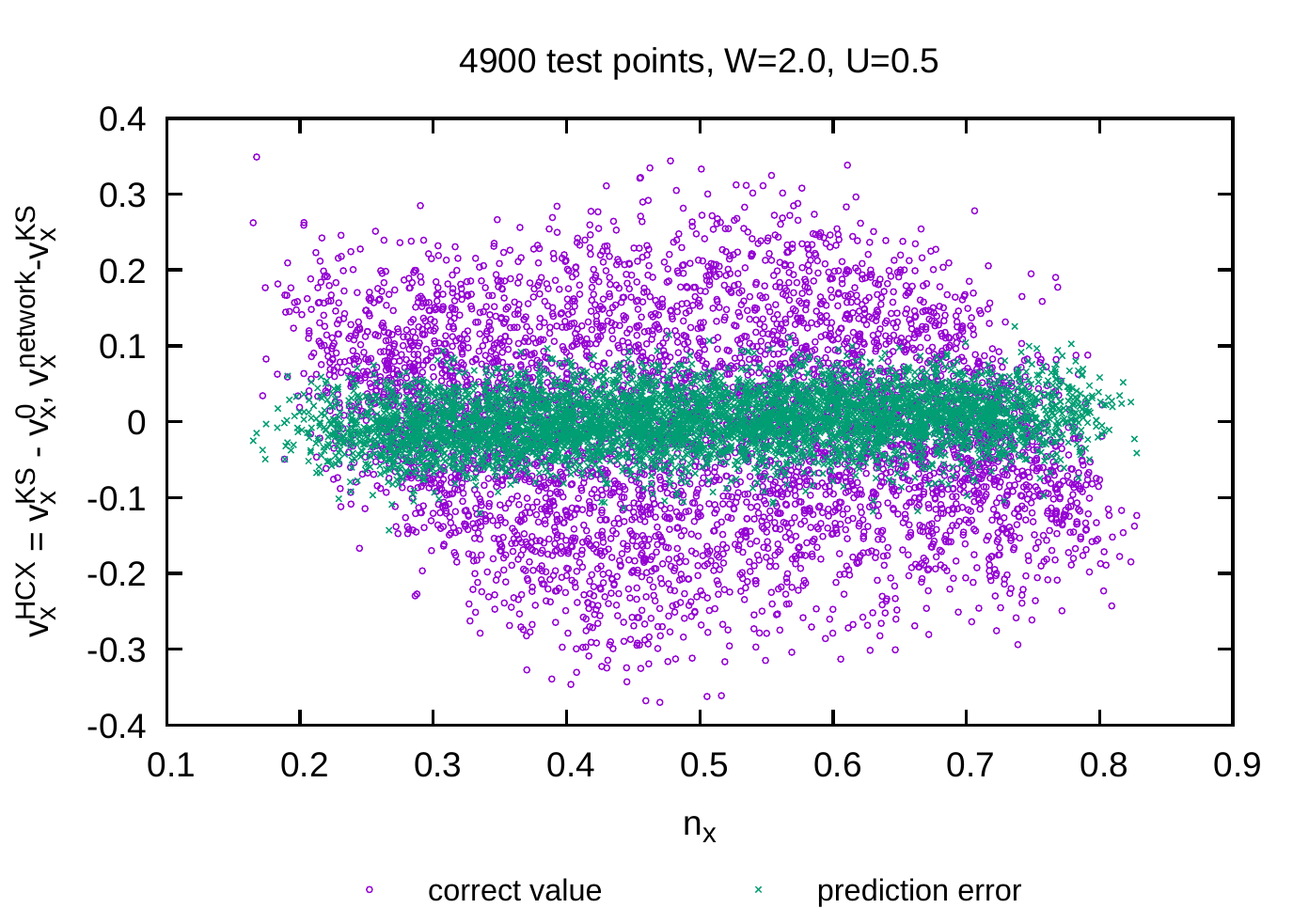}
\end{center}
\caption{\label{fig:Control} Results for the NN based DFT: (a) interaction induced $\vHXC$ for the test data vs.\ the local density;
(b) The $\vHXC$ potentials for 200 test values, the corresponding from the NN predicted values and arrows indicating the errors;
(c) the error arrows for 500 test values; (c) circles: the cloud of $\vHXC$ potentials as in Fig.~\ref{fig:Control}a, crosses: 
the error in the predicted Kohn-Sham values.}
\end{figure}

In Fig.~\ref{fig:Control}a we show the $\vHXC$ potentials vs.\ the local density.
Just by looking at the plot it is obvious that any local density approximation will  not be able
to describe the Kohn-Sham potentials. It is also hard to imagine that any density gradient based expansion
will be able to represent the rather dense cloud of potentials. The key point in local density approximations (LDA)
and gradient based expansions consists in assuming that the density functional is smooth  on short wave lengths and that
one can expand the functional around its local value. In contrast the  $\vHXC$ potentials as shown in Fig.~\ref{fig:Control}a
are highly oscillating and  multi valued. Even if one doesn't require a smooth density dependence of the  $\vHXC$ potentials  one
would need a least a strongly quenched distribution in order to construct an LDA based approximation.

Fig.~\ref{fig:Control}b compares the correct $\vHXC$ potentials with the ones predicted from the NN for the first 200 testing sets,
where the arrows denote the error of the prediction. In Fig.~\ref{fig:Control}c we display the same data, just drawing the
arrows for the errors for the first 500 testing sets. Here, for each sample  the error arrows start at the predicted value
and and at the desired correct value.
Finally in Fig.~\ref{fig:Control}d we show the cloud of $\vHXC$ potentials
for the complete testing set as in Fig.~\ref{fig:Control}a. In addition we also show the difference between the 
$\vKS$ as predicted from the NN minus the non-interaction potential. If the NN would work perfectly, those values
should zero. 

At a first sight the results looks rather disappointing. The NN is only capable to partially capture  the interaction induced
$\vHXC$ potentials. However, one should take into account that hand-crafting a density functional that reconstruct the cloud as
displayed Fig.~\ref{fig:Control}a is presumably impossible.

To make this result quantitative we show in Fig.~\ref{fig:Control_FSC}a the distribution function of the $\vHXC$ potentials
in comparison to the error of the predicted Kohn-Sham potentials, where both distributions can be fit by a Gaussian distribution.
The width of $\sigma=0.13$ for the $\vHXC$ is to be contrasted to the width of $\sigma=0.0347$ of the distribution of the
errors of the NN. Therefore, the NN is capable to capture about $73\%$ of the interaction effects.
While one would like to have a better agreement, the results shows that most of the correlations are already incorporated
and the NN might be able to provide correct trends.

One may try to improve the results by constructing more sophisticated NN. However, within this work we didn't succeed,
see Appendix~\ref{sec:CNN}.
We also note that it is straightforward to extract the total energy and the single particle gap from the inverse DFT
calculations, see \cite{Schmitteckert_Evers:PRL2008,PS:PCCP2018}, which could be incorporated into the output
of the NN.

\section{Up-scaling}
\label{sec:FSC}
\begin{figure}[ht]
\begin{center}
(a)\;\includegraphics[width=0.4\textwidth,valign=t]{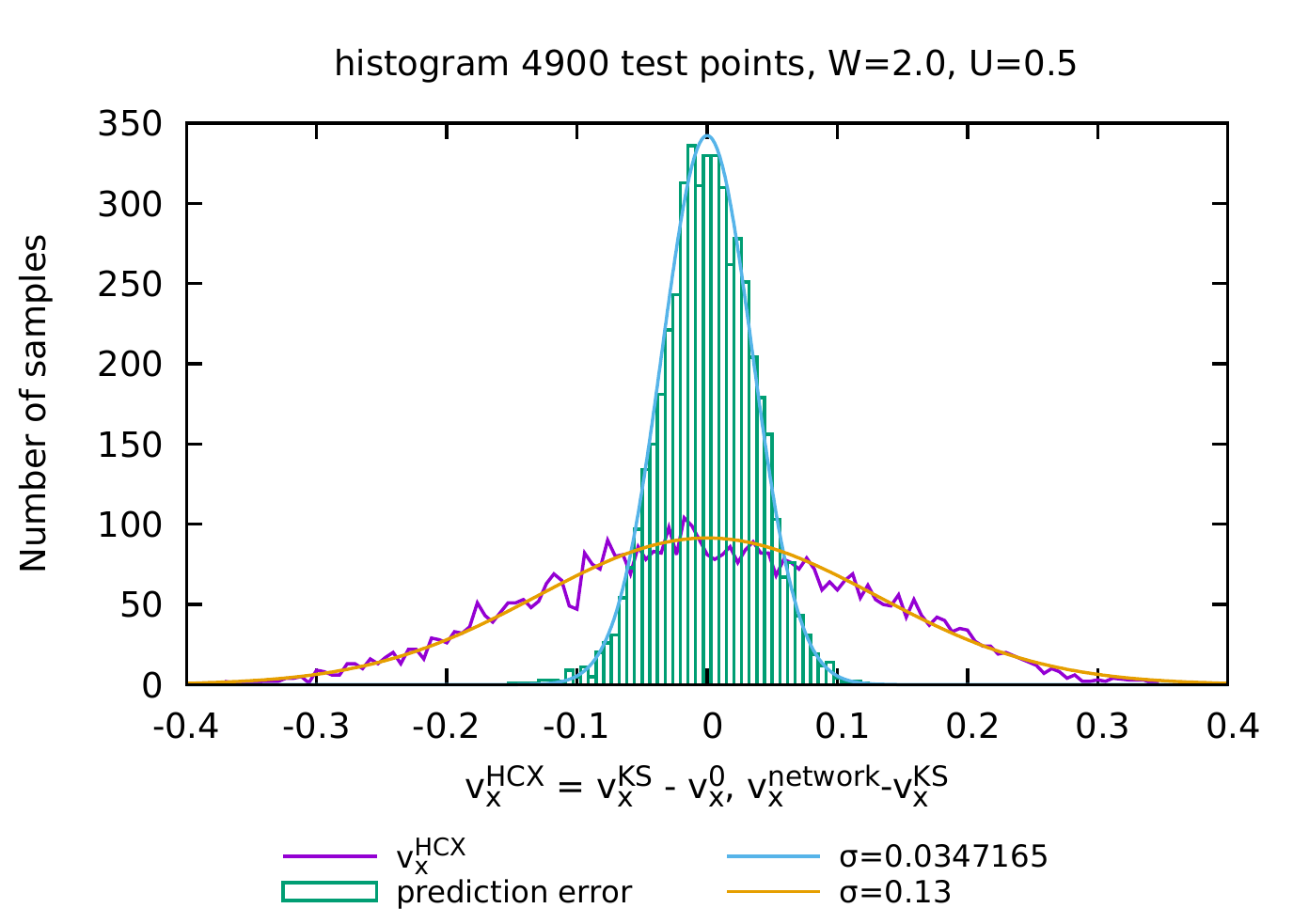}
\qquad
(b)\; \includegraphics[width=0.4\textwidth,valign=t]{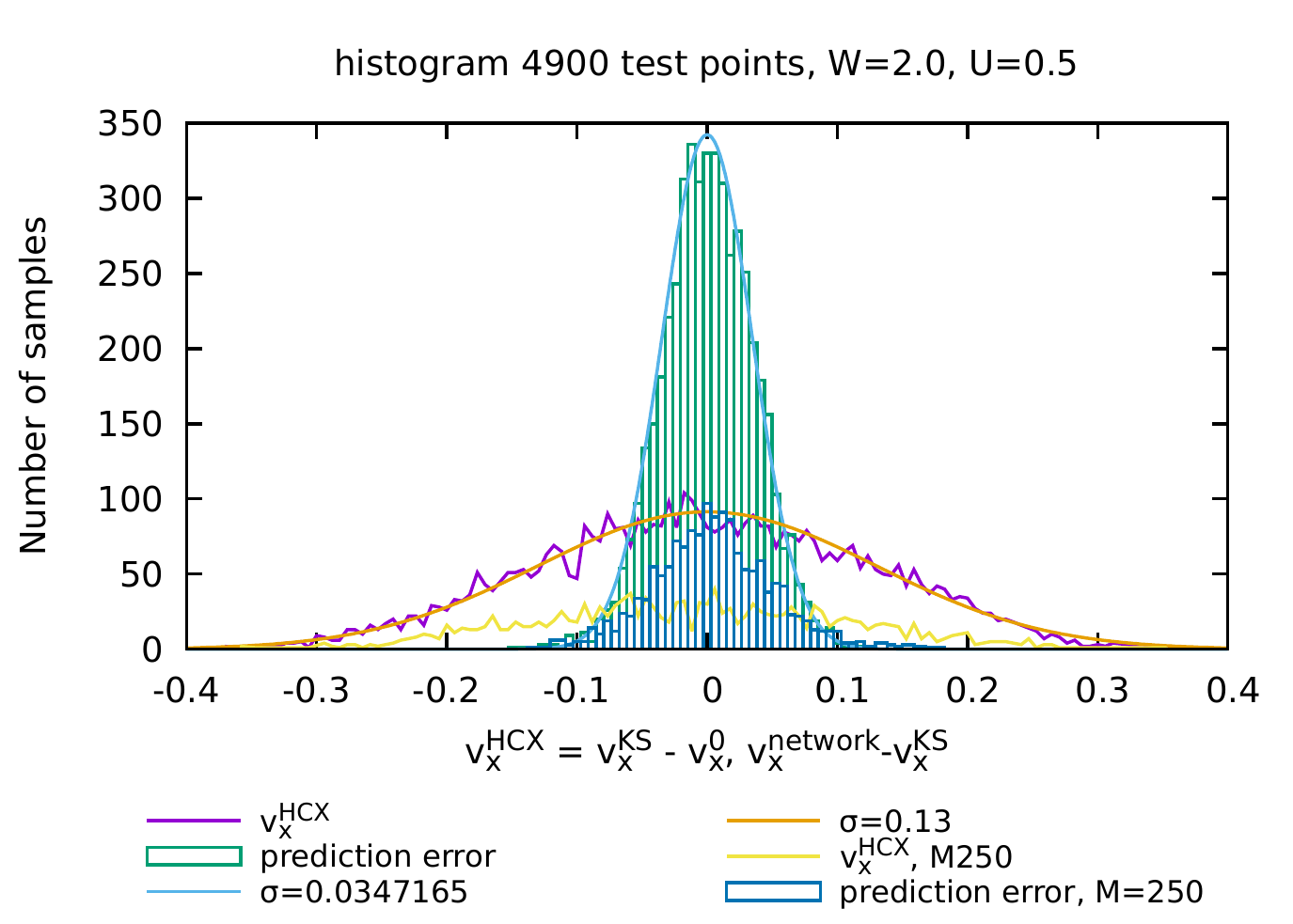}
\end{center}
\caption{\label{fig:Control_FSC} The error distribution of the NN: (a) The distribution of the 
$\vHXC$ potentials and the distribution of the error in the prediction of the $\vKS$ potentials
for the 98 site system and the corresponding Gaussian fits. 
(b) The same data as in (a) combined with the results obtained from up-scaling
the NN to 250 sites.}
\end{figure}

In this section we address the question whether we can use the approach to up-scale the calculations for larger system sizes,
i.e. using the NN trained from the $M=98$ site systems to obtain results for larger systems.
To this end we performed DMRG runs for ten $M=250$ site systems using the same DMRG parameter/setup as in Sec.~\ref{sec:NNDFT},
and extracted 2500 test sets as outlined in Eq.~\eqref{eq:set}.
In Fig.~\ref{fig:Control_FSC}b we compared the error of the Kohn-Sham potentials for the 250 site system obtained
from the NN trained with the $M=98$ sites data. 
Interestingly the width of the error distribution does not increase. The result therefore suggests
that the up-scaling of an NN trained on small systems to an evaluation on larger systems is
a fruitful concept.

\section{Summary}
In this work we applied the concept of deep learning via neural networks with the reverse engineering of
Kohn-Sham potentials in order to construct a DFT functional by learning the Kohn-Sham potentials.
We applied the idea to systems of one-dimensional
spinless fermions with nearest neighbor interaction and hopping combined with on-site disorder.
We showed, while being not perfect, we managed to capture $73\%$ of the interaction induced exchange correlation potentials.
In addition, we demonstrated that the concept of constructing functionals for the Kohn-Sham potentials from small system 
and to apply them for larger systems is a promising route for the investigation of interacting Fermi systems.

\begin{acknowledgements}
Most of the work reported here was performed while being at the university of W\"urzburg
and was supported by ERC-StG-Thomale-TOPOLECTRICS-336012 and was presented at the FQMT'19 in Praque.
We would like to thank Florian Eich for insightful discussions.

All authors contributed equally to the manuscript and the acquisition of the results.
\end{acknowledgements}

\bibliographystyle{unsrtnat}
\bibliography{references}

\begin{thebibliography}{38}
\providecommand{\natexlab}[1]{#1}
\providecommand{\url}[1]{\texttt{#1}}
\expandafter\ifx\csname urlstyle\endcsname\relax
  \providecommand{\doi}[1]{doi: #1}\else
  \providecommand{\doi}{doi: \begingroup \urlstyle{rm}\Url}\fi

\bibitem[White(1992)]{White:PRL92}
Steven~R. White.
\newblock Density matrix formulation for quantum renormalization groups.
\newblock \emph{Phys. Rev. Lett.}, 69:\penalty0 2863--2866, Nov 1992.
\newblock \doi{10.1103/PhysRevLett.69.2863}.
\newblock URL \url{http://link.aps.org/doi/10.1103/PhysRevLett.69.2863}.

\bibitem[White and Noack(1992)]{White_Noack:PRL92}
S.~R. White and R.~M. Noack.
\newblock Real-space quantum renormalization groups.
\newblock \emph{Phys. Rev. Lett.}, 68:\penalty0 3487--3490, Jun 1992.
\newblock \doi{10.1103/PhysRevLett.68.3487}.
\newblock URL \url{http://link.aps.org/doi/10.1103/PhysRevLett.68.3487}.

\bibitem[White(1993)]{White:PRB93}
Steven~R. White.
\newblock Density matrix renormalization group.
\newblock \emph{Phys.\ Rev.\ B}, 48:\penalty0 10345, 1993.

\bibitem[Noack and Manmana(2005)]{Noack_Manmana:AIPCP2005}
Reinhard~M. Noack and Salvatore~R. Manmana.
\newblock Diagonalization- and numerical renormalization-group-based methods
  for interacting quantum systems.
\newblock In Adolfo Avella and Ferdinando Mancini, editors, \emph{LECTURES ON
  THE PHYSICS OF HIGHLY CORRELATED ELECTRON SYSTEMS IX: Ninth Training Course
  in the Physics of Correlated Electron Systems and High-Tc Superconductors},
  volume 789, pages 93--163, Salerno, Italy, 2005. AIP.

\bibitem[Hallberg(2006)]{Hallberg:AP2006}
Karen~A. Hallberg.
\newblock New trends in density matrix renormalization.
\newblock \emph{Adv. Phys.}, 55\penalty0 (5):\penalty0 477--526, 2006.
\newblock \doi{http://dx.doi.org/10.1080/00018730600766432}.

\bibitem[Hohenberg and Kohn(1964)]{HohenbergKohn:1964}
P.~Hohenberg and W.~Kohn.
\newblock Inhomogeneous electron gas.
\newblock \emph{Phys. Rev.}, 136:\penalty0 B864--B871, Nov 1964.
\newblock \doi{10.1103/PhysRev.136.B864}.
\newblock URL \url{https://link.aps.org/doi/10.1103/PhysRev.136.B864}.

\bibitem[Kohn and Sham(1965)]{KohnSham:1965}
W.~Kohn and L.~J. Sham.
\newblock Self-consistent equations including exchange and correlation effects.
\newblock \emph{Phys. Rev.}, 140:\penalty0 A1133--A1138, Nov 1965.
\newblock \doi{10.1103/PhysRev.140.A1133}.
\newblock URL \url{https://link.aps.org/doi/10.1103/PhysRev.140.A1133}.

\bibitem[Dreizler and Gross(1990)]{dreizler90}
R.M. Dreizler and E.K.U. Gross.
\newblock \emph{Density Functional Theory}.
\newblock Springer--Verlag, 1990.

\bibitem[Gunnarsson and Sch\"onhammer(1986)]{SchoenhammerGunnarsson:PRL1986}
O.~Gunnarsson and K.~Sch\"onhammer.
\newblock Density-functional treatment of an exactly solvable semiconductor
  model.
\newblock \emph{Phys. Rev. Lett.}, 56\penalty0 (18):\penalty0 1968--1971, May
  1986.
\newblock \doi{10.1103/PhysRevLett.56.1968}.

\bibitem[Schmitteckert and Evers(2008)]{Schmitteckert_Evers:PRL2008}
Peter Schmitteckert and Ferdinand Evers.
\newblock Exact ground state density-functional theory for impurity models
  coupled to external reservoirs and transport calculations.
\newblock \emph{Phys. Rev. Lett.}, 100\penalty0 (8):\penalty0 086401, Feb 2008.
\newblock \doi{10.1103/PhysRevLett.100.086401}.

\bibitem[Schmitteckert(2018)]{PS:PCCP2018}
Peter Schmitteckert.
\newblock Inverse mean field theories.
\newblock \emph{Phys. Chem. Chem. Phys.}, 20:\penalty0 27600--27610, 2018.
\newblock \doi{10.1039/C8CP03763A}.
\newblock URL \url{http://dx.doi.org/10.1039/C8CP03763A}.

\bibitem[Nielsen(2015)]{Nielsen:DP2015}
Michael~A. Nielsen.
\newblock \emph{Neural Networks and Deep Learning}.
\newblock Determination Press, 2015.

\bibitem[McCulloch and Pitts(1943)]{McCulloch:BMB1943}
Warren~S. McCulloch and Walter Pitts.
\newblock A logical calculus of the ideas immanent in nervous activity.
\newblock \emph{The bulletin of mathematical biophysics}, 5\penalty0
  (4):\penalty0 115--133, Dec 1943.
\newblock ISSN 1522-9602.
\newblock \doi{10.1007/BF02478259}.
\newblock URL \url{https://doi.org/10.1007/BF02478259}.

\bibitem[Rumelhart et~al.(1986)Rumelhart, E., and
  William]{Rumelhart:Nature1986}
D.~E. Rumelhart, Hinton~G. E., and R.~J. William.
\newblock Learning represantations by back-propagating errors.
\newblock \emph{Nature}, 323:\penalty0 533, 1986.

\bibitem[LeCun and Bengio(1995)]{LeCun:1995}
Y~LeCun and Y~Bengio.
\newblock Convolutional networks for images, speech, and time series.
\newblock \emph{The handbook of brain theory and neural networks},
  3361:\penalty0 3539, 1995.

\bibitem[Nomi and {et.al.}()]{tinyDNN}
Taiga Nomi and {et.al.}
\newblock tiny dnn.

\bibitem[Abadi et~al.(2015)Abadi, Agarwal, Barham, Brevdo, Chen, Citro,
  Corrado, Davis, Dean, Devin, Ghemawat, Goodfellow, Harp, Irving, Isard, Jia,
  Jozefowicz, Kaiser, Kudlur, Levenberg, Man\'{e}, Monga, Moore, Murray, Olah,
  Schuster, Shlens, Steiner, Sutskever, Talwar, Tucker, Vanhoucke, Vasudevan,
  Vi\'{e}gas, Vinyals, Warden, Wattenberg, Wicke, Yu, and
  Zheng]{tensorflow:2015}
Mart\'{\i}n Abadi, Ashish Agarwal, Paul Barham, Eugene Brevdo, Zhifeng Chen,
  Craig Citro, Greg~S. Corrado, Andy Davis, Jeffrey Dean, Matthieu Devin,
  Sanjay Ghemawat, Ian Goodfellow, Andrew Harp, Geoffrey Irving, Michael Isard,
  Yangqing Jia, Rafal Jozefowicz, Lukasz Kaiser, Manjunath Kudlur, Josh
  Levenberg, Dandelion Man\'{e}, Rajat Monga, Sherry Moore, Derek Murray, Chris
  Olah, Mike Schuster, Jonathon Shlens, Benoit Steiner, Ilya Sutskever, Kunal
  Talwar, Paul Tucker, Vincent Vanhoucke, Vijay Vasudevan, Fernanda Vi\'{e}gas,
  Oriol Vinyals, Pete Warden, Martin Wattenberg, Martin Wicke, Yuan Yu, and
  Xiaoqiang Zheng.
\newblock {TensorFlow}: Large-scale machine learning on heterogeneous systems,
  2015.
\newblock URL \url{https://www.tensorflow.org/}.
\newblock Software available from tensorflow.org.

\bibitem[Chollet et~al.(2015)]{keras:2015}
Francois Chollet et~al.
\newblock Keras, 2015.

\bibitem[Carleo et~al.(2019)Carleo, Cirac, Cranmer, Daudet, Schuld, Tishby,
  Vogt-Maranto, and Zdeborov\'a]{RevModPhys.91.045002}
Giuseppe Carleo, Ignacio Cirac, Kyle Cranmer, Laurent Daudet, Maria Schuld,
  Naftali Tishby, Leslie Vogt-Maranto, and Lenka Zdeborov\'a.
\newblock Machine learning and the physical sciences.
\newblock \emph{Rev. Mod. Phys.}, 91:\penalty0 045002, Dec 2019.
\newblock \doi{10.1103/RevModPhys.91.045002}.
\newblock URL \url{https://link.aps.org/doi/10.1103/RevModPhys.91.045002}.

\bibitem[Brockherde et~al.(2017)Brockherde, Vogt, Li, Tuckerman, Burke, and
  M{\"u}ller]{Brockherde:NC2017}
Felix Brockherde, Leslie Vogt, Li~Li, Mark~E. Tuckerman, Kieron Burke, and
  Klaus-Robert M{\"u}ller.
\newblock Bypassing the kohn-sham equations with machine learning.
\newblock \emph{Nature Communications}, 8\penalty0 (1):\penalty0 872, Oct 2017.
\newblock ISSN 2041-1723.
\newblock \doi{10.1038/s41467-017-00839-3}.
\newblock URL \url{https://doi.org/10.1038/s41467-017-00839-3}.

\bibitem[Kolb et~al.(2017)Kolb, Lentz, and Kolpak]{Kolb:SR2017}
Brian Kolb, Levi~C. Lentz, and Alexie~M. Kolpak.
\newblock Discovering charge density functionals and structure-property
  relationships with prophet: A general framework for coupling machine learning
  and first-principles methods.
\newblock \emph{Scientific Reports}, 7\penalty0 (1):\penalty0 1192, Apr 2017.
\newblock ISSN 2045-2322.
\newblock \doi{10.1038/s41598-017-01251-z}.
\newblock URL \url{https://doi.org/10.1038/s41598-017-01251-z}.

\bibitem[Hu et~al.(2003)Hu, Wang, Wong, and Chen]{Hu:JCP2003}
L.~Hu, X.~Wang, L.~Wong, and G.~Chen.
\newblock Combined first-principles calculation and neural-network correction
  approach for heat of formation.
\newblock \emph{J. Chem. Phys.}, 119:\penalty0 11501, 2003.

\bibitem[Zheng et~al.(2004)Zheng, Hu, Wang, and Chen]{ZHENG:CPL2004}
Xiao Zheng, LiHong Hu, XiuJun Wang, and GuanHua Chen.
\newblock A generalized exchange-correlation functional: the neural-networks
  approach.
\newblock \emph{Chemical Physics Letters}, 390\penalty0 (1):\penalty0 186 --
  192, 2004.
\newblock ISSN 0009-2614.
\newblock \doi{https://doi.org/10.1016/j.cplett.2004.04.020}.
\newblock URL
  \url{http://www.sciencedirect.com/science/article/pii/S0009261404005603}.

\bibitem[Liu et~al.(2017)Liu, Wang, Du, Hu, Zheng, and Chen]{Liu:JPCA2017}
Qin Liu, JingChun Wang, PengLi Du, LiHong Hu, Xiao Zheng, and GuanHua Chen.
\newblock Improving the performance of long-range-corrected
  exchange-correlation functional with an embedded neural network.
\newblock \emph{The Journal of Physical Chemistry A}, 121\penalty0
  (38):\penalty0 7273--7281, 2017.
\newblock \doi{10.1021/acs.jpca.7b07045}.
\newblock URL \url{ttps://doi.org/10.1021/acs.jpca.7b07045}.
\newblock PMID: 28876064.

\bibitem[Snyder et~al.(2012)Snyder, Rupp, Hansen, M\"uller, and
  Burke]{Snyder:PRL2012}
John~C. Snyder, Matthias Rupp, Katja Hansen, Klaus-Robert M\"uller, and Kieron
  Burke.
\newblock Finding density functionals with machine learning.
\newblock \emph{Phys. Rev. Lett.}, 108:\penalty0 253002, Jun 2012.
\newblock \doi{10.1103/PhysRevLett.108.253002}.
\newblock URL \url{https://link.aps.org/doi/10.1103/PhysRevLett.108.253002}.

\bibitem[Snyder et~al.(2013)Snyder, Rupp, Hansen, Blooston, Müller, and
  Burke]{Snyder:JCP2013}
John~C. Snyder, Matthias Rupp, Katja Hansen, Leo Blooston, Klaus-Robert
  Müller, and Kieron Burke.
\newblock Orbital-free bond breaking via machine learning.
\newblock \emph{The Journal of Chemical Physics}, 139\penalty0 (22):\penalty0
  224104, 2013.
\newblock \doi{10.1063/1.4834075}.
\newblock URL \url{https://doi.org/10.1063/1.4834075}.

\bibitem[Li et~al.(2016)Li, Baker, White, and Burke]{Burke:PRB2016}
Li~Li, Thomas~E. Baker, Steven~R. White, and Kieron Burke.
\newblock Pure density functional for strong correlation and the thermodynamic
  limit from machine learning.
\newblock \emph{Phys. Rev. B}, 94:\penalty0 245129, Dec 2016.
\newblock \doi{10.1103/PhysRevB.94.245129}.
\newblock URL \url{https://link.aps.org/doi/10.1103/PhysRevB.94.245129}.

\bibitem[Giamarchi and Schulz(1988)]{GiamarchiSchulz:PRB1988}
T.~Giamarchi and H.~J. Schulz.
\newblock Anderson localization and interactions in one-dimensional metals.
\newblock \emph{Phys. Rev. B}, 37:\penalty0 325--340, Jan 1988.
\newblock \doi{10.1103/PhysRevB.37.325}.
\newblock URL \url{https://link.aps.org/doi/10.1103/PhysRevB.37.325}.

\bibitem[Schmitteckert et~al.(1998{\natexlab{a}})Schmitteckert, Schulze,
  Schuster, Schwab, and Eckern]{PS:PRL1998}
P.~Schmitteckert, T.~Schulze, C.~Schuster, P.~Schwab, and U.~Eckern.
\newblock Anderson localization versus delocalization of interacting fermions
  in one dimension.
\newblock \emph{Phys. Rev. Lett.}, 80:\penalty0 560--563, Jan
  1998{\natexlab{a}}.
\newblock \doi{10.1103/PhysRevLett.80.560}.
\newblock URL \url{https://link.aps.org/doi/10.1103/PhysRevLett.80.560}.

\bibitem[Schmitteckert et~al.(1998{\natexlab{b}})Schmitteckert, Jalabert,
  Weinmann, and Pichard]{PS:PRL1998b}
Peter Schmitteckert, Rodolfo~A. Jalabert, Dietmar Weinmann, and Jean-Louis
  Pichard.
\newblock From the fermi glass towards the mott insulator in one dimension:
  Delocalization and strongly enhanced persistent currents.
\newblock \emph{Phys. Rev. Lett.}, 81:\penalty0 2308--2311, Sep
  1998{\natexlab{b}}.
\newblock \doi{10.1103/PhysRevLett.81.2308}.
\newblock URL \url{https://link.aps.org/doi/10.1103/PhysRevLett.81.2308}.

\bibitem[Jalabert et~al.(2001)Jalabert, Weinmann, and Pichard]{Jalabert:PE2001}
Rodolfo~A Jalabert, Dietmar Weinmann, and Jean-Louis Pichard.
\newblock Partial delocalization of the ground state by repulsive interactions
  in a disordered chain.
\newblock \emph{Physica E: Low-dimensional Systems and Nanostructures},
  9\penalty0 (3):\penalty0 347 -- 351, 2001.
\newblock ISSN 1386-9477.
\newblock \doi{https://doi.org/10.1016/S1386-9477(00)00226-5}.
\newblock URL
  \url{http://www.sciencedirect.com/science/article/pii/S1386947700002265}.
\newblock Proceedings of an International Workshop and Seminar on the Dynamics
  of Complex Systems.

\bibitem[Schmitteckert(1999)]{PS:Proceedings98}
Peter Schmitteckert.
\newblock Disordered one-dimensional fermi systems.
\newblock In \emph{Density Matrix Renormalization\cite{Proceedings98}}, pages
  345--355, 1999.
\newblock ISBN 978-3-540-66129-0.

\bibitem[Peschel et~al.(1999)Peschel, Wang, M.Kaulke, and
  Hallberg]{Proceedings98}
I.~Peschel, X.~Wang, M.Kaulke, and K.~Hallberg, editors.
\newblock \emph{Density Matrix Renormalization}, 1999.
\newblock ISBN 978-3-540-66129-0.

\bibitem[Sch\"onhammer et~al.(1995)Sch\"onhammer, Gunnarsson, and
  Noack]{SchoenhammerGunnarssonNoack:PRB1995}
K.~Sch\"onhammer, O.~Gunnarsson, and R.~M. Noack.
\newblock Density-functional theory on a lattice: Comparison with exact
  numerical results for a model with strongly correlated electrons.
\newblock \emph{Phys. Rev. B}, 52:\penalty0 2504--2510, Jul 1995.
\newblock \doi{10.1103/PhysRevB.52.2504}.
\newblock URL \url{https://link.aps.org/doi/10.1103/PhysRevB.52.2504}.

\bibitem[Evers and Schmitteckert(2013)]{FE_PS:PSS2013}
Ferdinand Evers and Peter Schmitteckert.
\newblock Density functional theory with exact xc-potentials: Lessons from
  dmrg-studies and exactly solvable models.
\newblock \emph{Phys. Status Solidi B}, 250:\penalty0 2330, 2013.

\bibitem[Robbins and Monro(1951)]{robbins:AMS1951}
Herbert Robbins and Sutton Monro.
\newblock A stochastic approximation method.
\newblock \emph{Ann. Math. Statist.}, 22\penalty0 (3):\penalty0 400--407, 09
  1951.
\newblock \doi{10.1214/aoms/1177729586}.
\newblock URL \url{https://doi.org/10.1214/aoms/1177729586}.

\bibitem[Kingma and Ba(2014)]{kingma2014adam}
Diederik~P. Kingma and Jimmy Ba.
\newblock Adam: A method for stochastic optimization, 2014.

\bibitem[Duchi et~al.(2011)Duchi, Hazan, and Singer]{Duchi:JMLR2011}
John Duchi, Elad Hazan, and Yoram Singer.
\newblock Adaptive subgradient methods for online learning and stochastic
  optimization.
\newblock \emph{JMLR}, 12:\penalty0 2121--2159, 2011.

\end{thebibliography}

\appendix

\section{Neural networks for fitting functions}
\label{sec:FitNN}
The main application of neural networks consists in the classification of input variables, i.e. one maps the input to a discrete set
of output variable, with the standard internet example of {\em``is it a cat or not?''}. 
Here we provide an example for applying a neural network on fitting a function.

\subsection{The network}
\begin{figure}[ht]
\begin{center}
(a)\; \includegraphics[width=0.3\textwidth,valign=m]{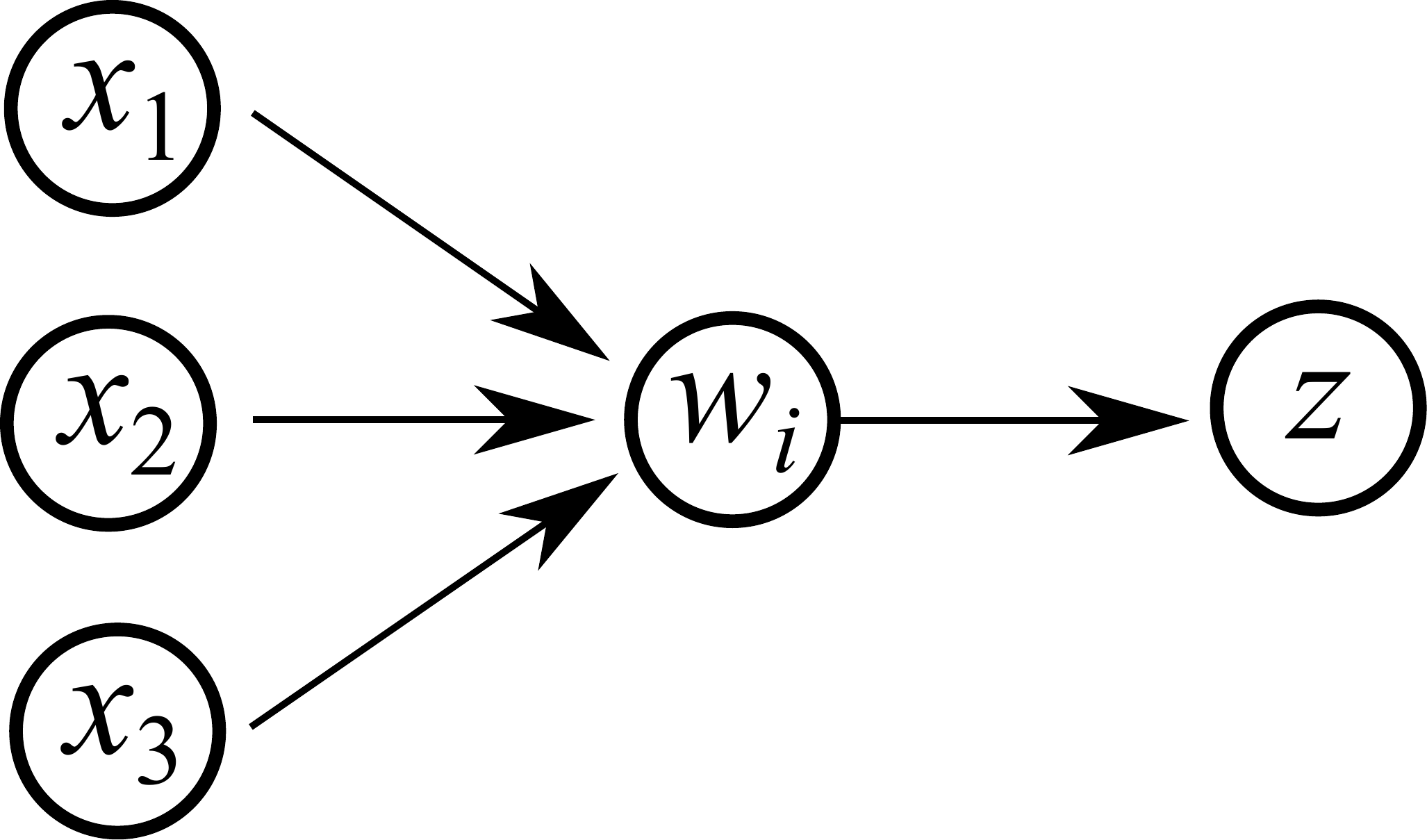}
\qquad
(b)\; \includegraphics[width=0.5\textwidth,valign=m]{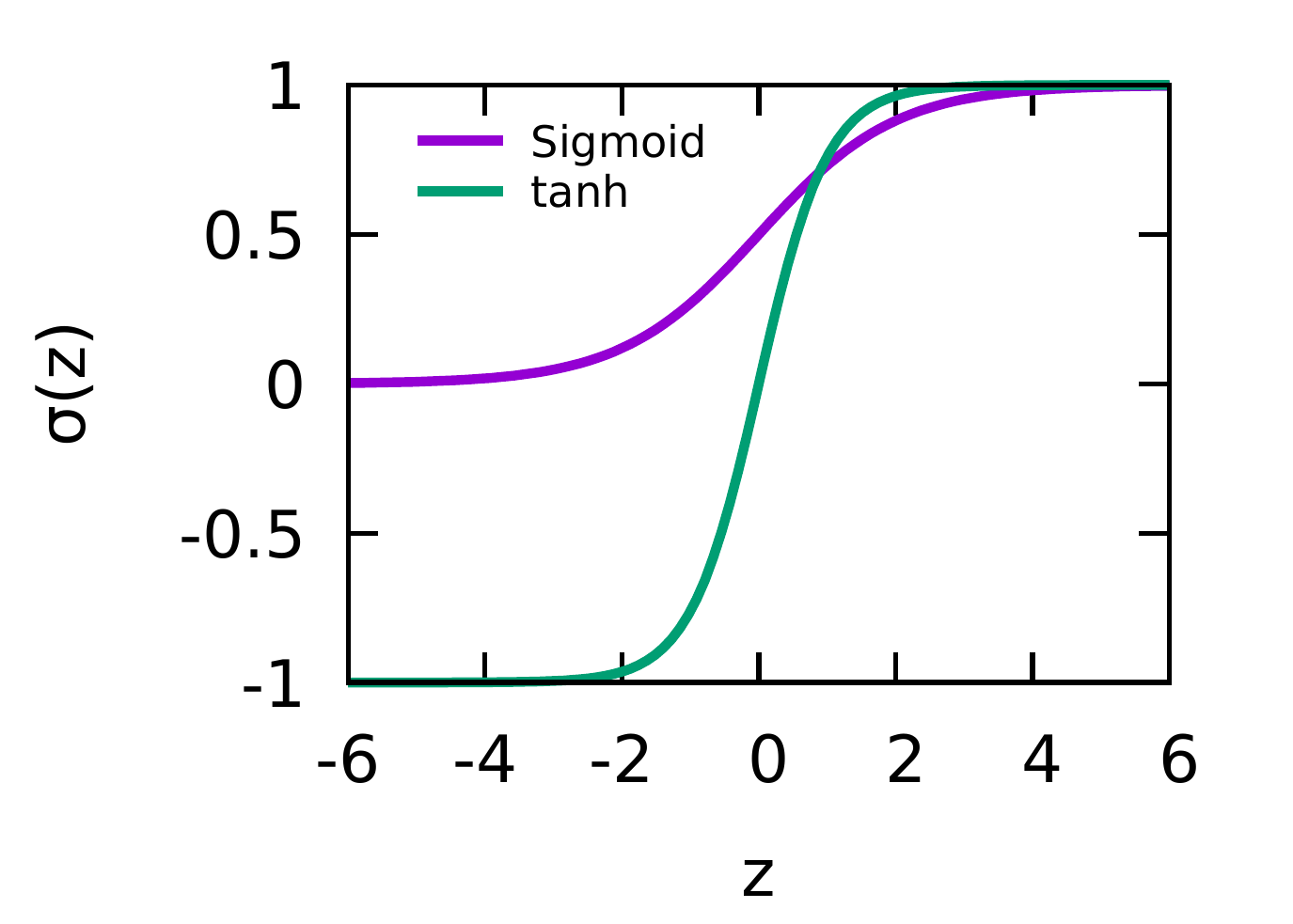}
\end{center}
\caption{\label{fig:NN-neuron} The building block of a NN: (a) a neutron (b) typical weight function: a sigmoid and a $\tanh$.}
\end{figure}

The basic building block of a neural network (NN) consists of a neutron as depicted in Fig.~\ref{fig:NN-neuron}a.

The neuron consist of an input $\{x_i\}$, weight factors $\{w_i\}$, an offset $b$, 
a so-called activation function $\sigma(z)$, see Fig.~\ref{fig:NN-neuron}b,
and the output $z$:
\begin{equation}
 z \;=\; \sigma\left( b + \sum_\ell w_\ell x_\ell \right) \label{eq:ParameterNeuron}
\end{equation}
Throughout this work we have always used an $\tanh$ activation function.
\begin{figure}[ht]
\begin{center}
  \includegraphics[width=0.8\textwidth]{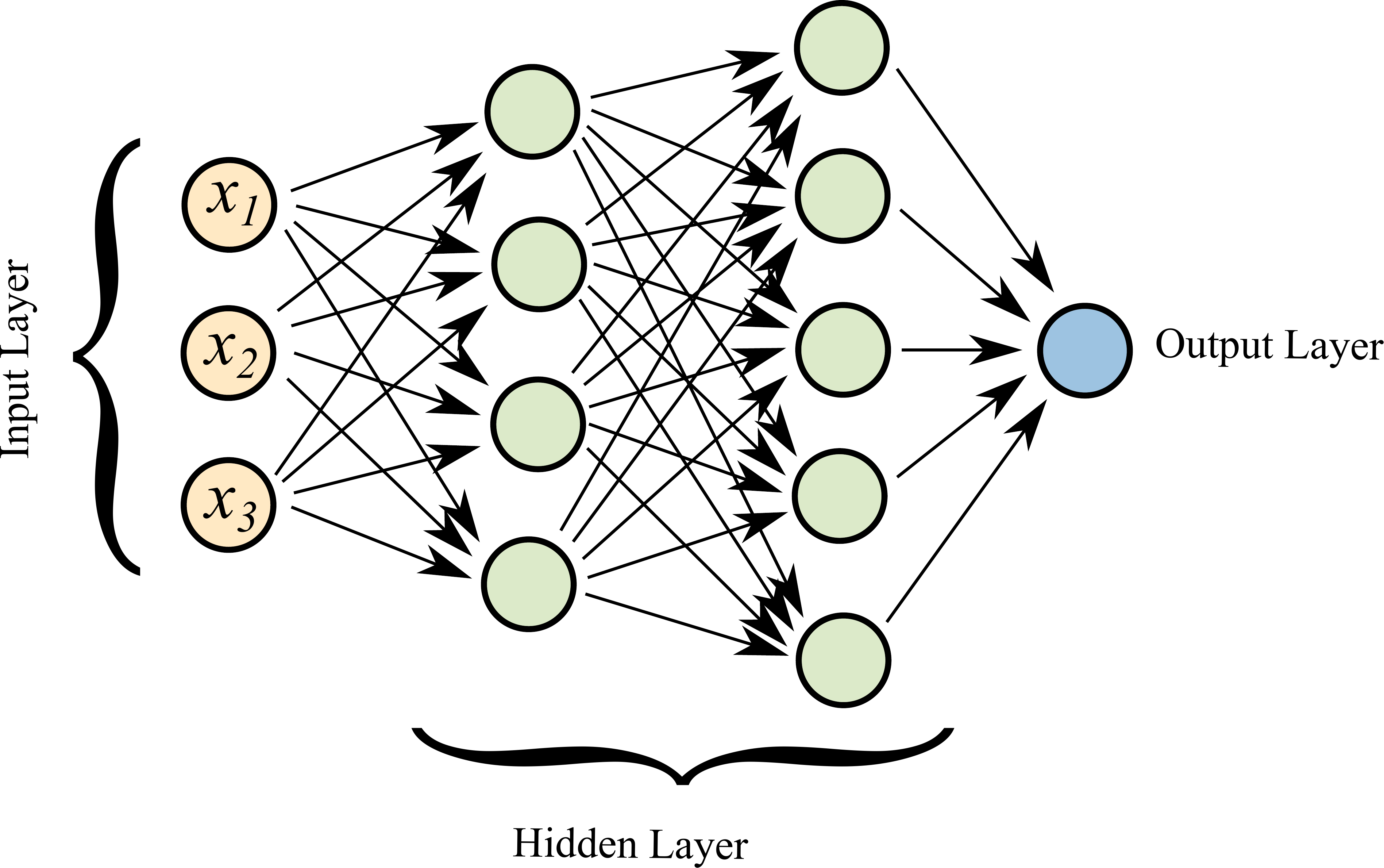}
\end{center}
\caption{\label{fig:NN-network} A neural network build out of neurons shown in Fig.~\ref{fig:NN-neuron}a.}
\end{figure}
One now combines many neurons, Fig.~\ref{fig:NN-neuron}a, into a neural network in a layered fashion 
by connecting inputs of the neurons of one layer with the outputs of the neurons of the previous layer, see Fig.~\ref{fig:NN-network}.
Since from a user perspective the NN in Fig.~\ref{fig:NN-network} translates the input of the first layer into the output of the last layer
one calls the first layer the {\em input} layer, the last one the {\em output} layer and the other layers are denoted as {\em hidden} layers.
If each neuron is connected to each neuron of the previous layer one calls the network  {\em dense}.
The training of a NN is often referred to as machine learning, and in the presence of many hidden layer
as deep learning.

In summary, the NN in  Fig.~\ref{fig:NN-network} calculates an output $z$ from the input $\{x_j\}$, where one has to specify 
the parameter in Eq.~\eqref{eq:ParameterNeuron} for each neuron $n$ in layer $\ell$,
\begin{equation}
 z_{\ell,n} \;=\; \sigma\left( b_{\ell,n} + \sum_j w_{\ell,n,j} x_j \right) \,. \label{eq:ParameterNetwork}
\end{equation}
Of course, this can be extended to create multiple output variables $z_k$ in the output layer.

In order to apply a NN for fitting functions $f(x)$ we use a NN with a single input $x$ and a single output  $z$.
The free parameter $\{b_{\ell,n},  w_{\ell,n,j}\}$ are the set of fitting parameter. We would like to note that this approach
is in contrast to the desired approach in physics, where on tries to fit a phenomenon with a suitable function using as few fitting parameter
as possible. Instead, here we take the opposite approach by using a simple fitting function unrelated to the problem and fit the desired
function with a large number of parameter and a few steps of recursion.

\subsection{Training the neural network: Minimize Cost Function}
The idea to determine the fit parameter for fitting a function $f({\bf x})$ consists in minimizing a cost function, typically
\begin{equation}
 C({\bf x}) \;=\; \frac{1}{N} \sum_{i=1}^{N} \left|\left| f({\bf x}_i) - z_{\{b_{\ell,n}, w_{\ell,n,j}\}}({\bf x}_i) \right|\right|_2  \label{eq:cost}
\end{equation}
where $N$ denotes the number of training samples.
Eq.~\eqref{eq:cost} could in principle be minimized by a standard steepest descent gradient search.
However, due the vast amount of fit parameter this is not feasible in non-trivial examples, as the number of parameter,
and therefore the dimensions of the associated matrices get too large.
The breakthrough for neural networks was provided by the invention of the 
back propagating algorithm \cite{Rumelhart:Nature1986}
combined with a stochastic evaluation of the gradients \cite{robbins:AMS1951,kingma2014adam,Duchi:JMLR2011}
combined the massive computational power
of graphic cards, and for pattern recognition the use of convolutional layers \cite{LeCun:1995}, see below.
In the example provided in this section we used {\tt tensorflow} \cite{tensorflow:2015} software package 
combined with the {\tt keras} \cite{keras:2015} front end.

\subsection{An example}
\label{sec:example}
As an example we look at the function
\begin{equation}
 f(x) \;=\ \sin(3x) \,-\, 0.8 \cos^2(13x) \e^{0.5 x} \,+\, 4 \e^{-\frac{x^2}{0.0005}}  \,+\, 6 \e^{ -\frac{(x+0.4)^2}{0.0001}} \label{eq:f}
\end{equation}
which has no deeper meaning, it was just handcrafted to represent a not too trivial function combining  sharp and non-sharp features.

\begin{figure}
\begin{center}
(a) \;\includegraphics[width=0.4\textwidth,valign=t]{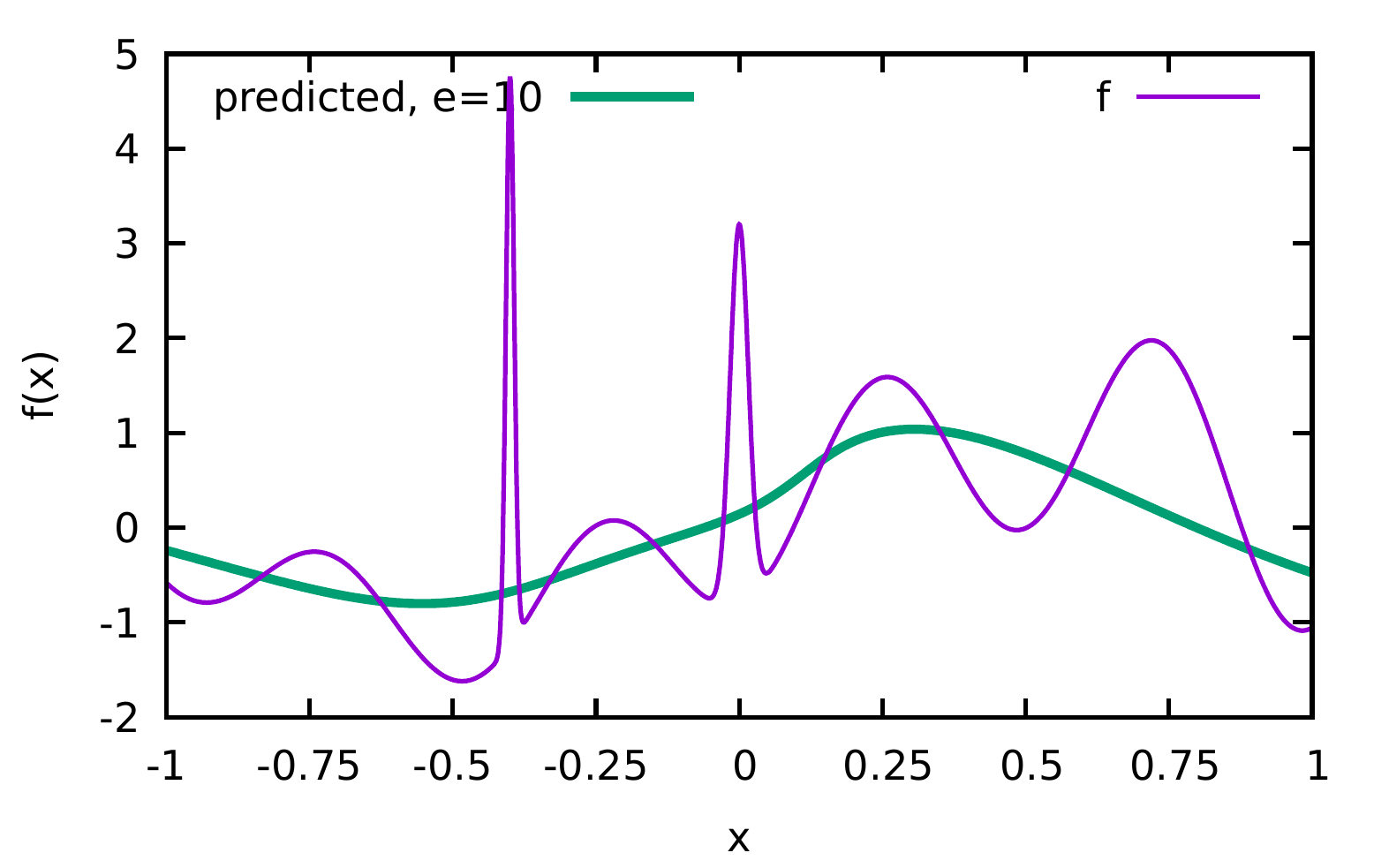} \qquad
(b) \;\includegraphics[width=0.4\textwidth,valign=t]{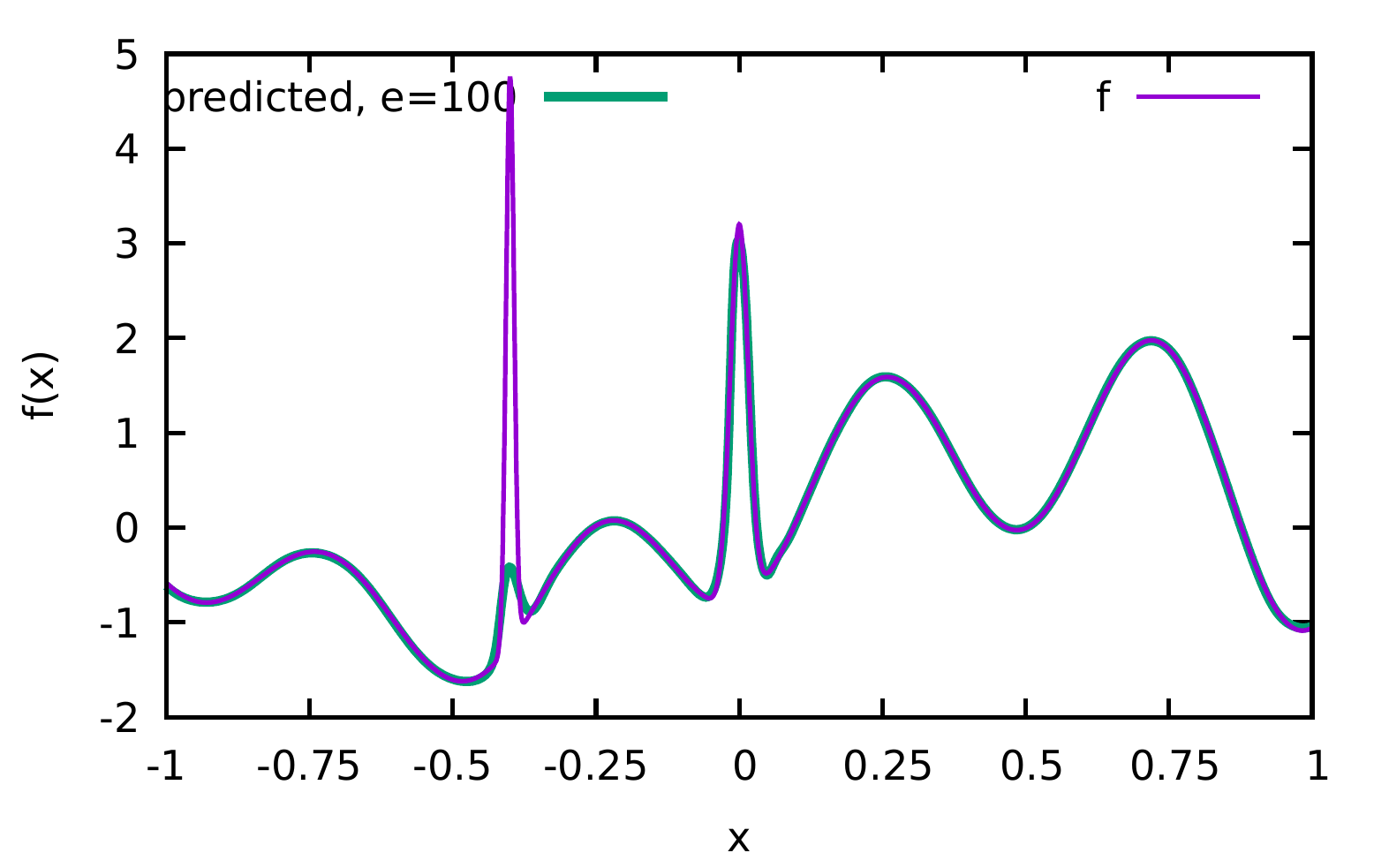} \\
(c) \;\includegraphics[width=0.4\textwidth,valign=t]{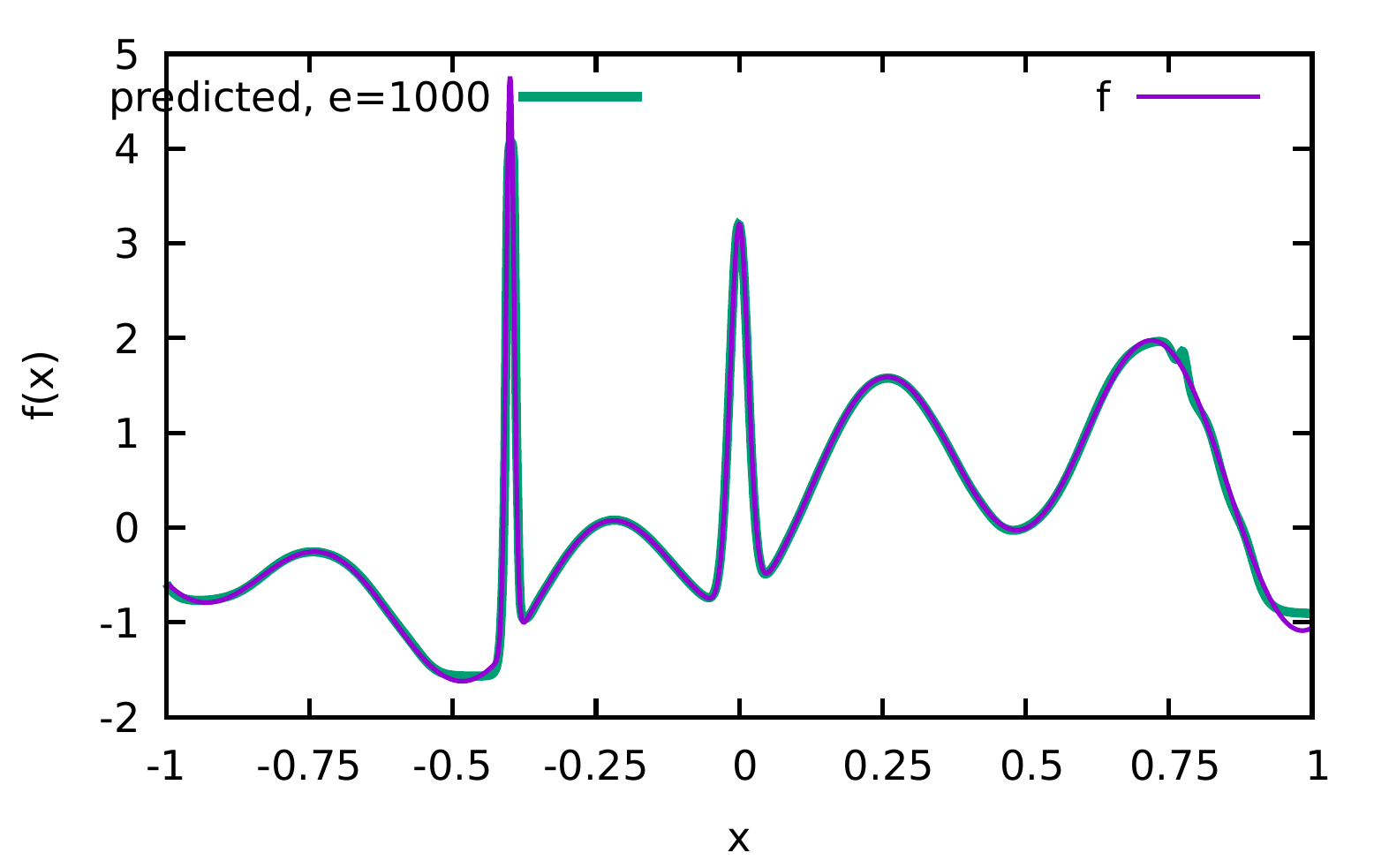} \qquad
(d) \;\includegraphics[width=0.4\textwidth,valign=t]{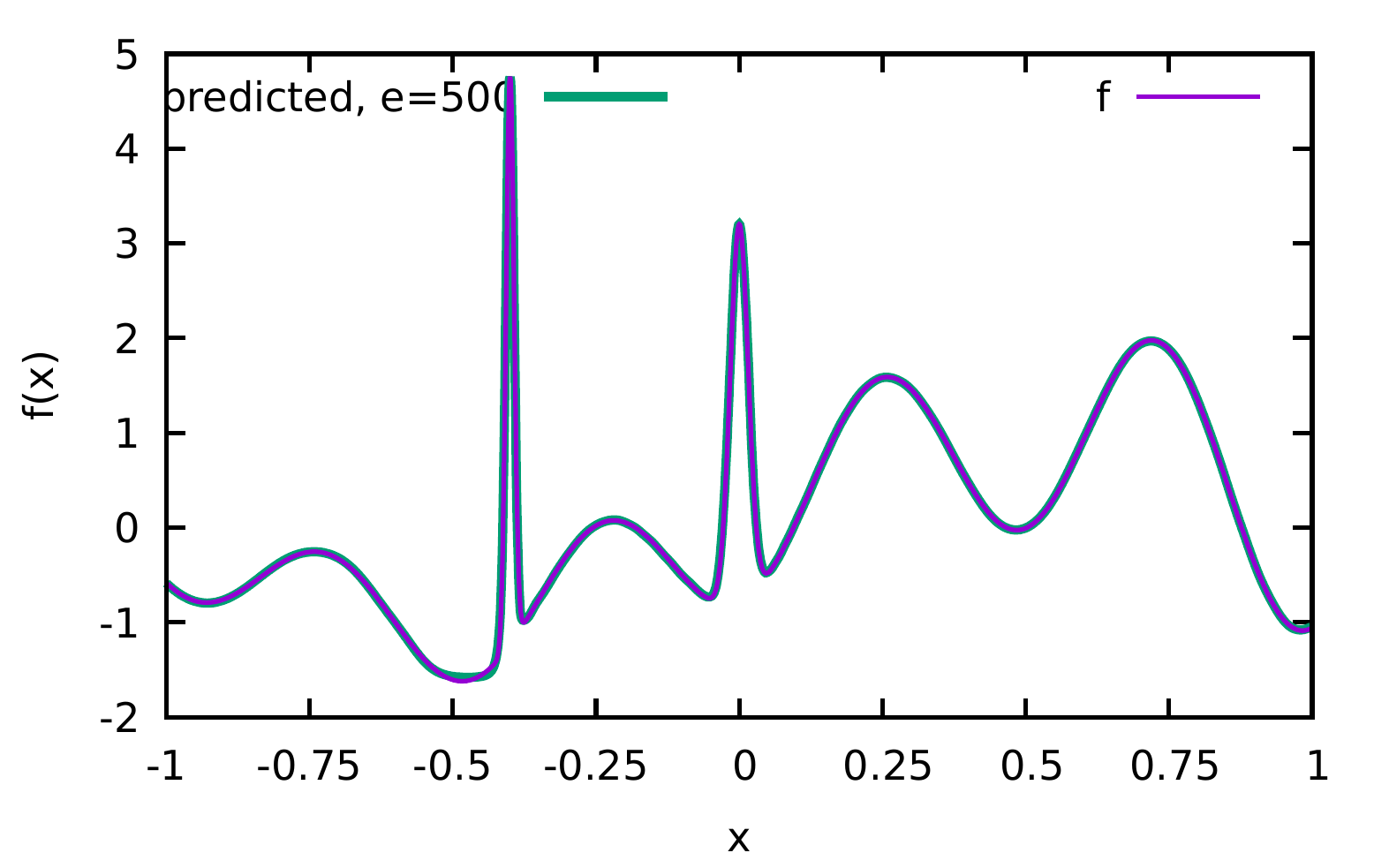} 
\end{center}
\caption{\label{fig:Fit_function50} Fits for the function $f(x)$ Eq.~\eqref{eq:f} using a $\tanh$ activation function obtained
 using {\tt tensorflow/keras} for a $1 \times 50 \times 50 \times 1$ network.
 (a)  25.000 samples, 10 repetitions of SGD;
 (b)  25.000 samples, 100 repetitions of SGD;
 (c)  25.000 samples, 1000 repetitions of SGD;
 (d)  $2\times$ 25.000 samples, 250 repetitions ($1\times$ SGD \&  $1\times$ADAM). 
}
\end{figure}

Since $f$ is a single valued single argument function the input and output layer consists of a single neuron only.
In Fig.~\ref{fig:Fit_function50} we show the results for fitting the function $f(x)$ in Eq.~\eqref{eq:f} with two hidden 
layers consisting of fifty neurons each. In result we applied a dense NN with a $1 \times 50 \times 50 \times 1$ structure.
In order to train the system we generated 25.000 random values $x_j$ with the corresponding $z_j = f(x_j)$.
We then trained the NN by performing a stochastic gradient descent search (SGD) with ten repetitions over the complete set of $\{x_j, z_j\}$.
We then evaluate the NN on an equidistantly spaced set of   $\{x_\ell\}$. As one can see in Fig.~\ref{fig:Fit_function50}a the
result is a rather smooth function that misses the sharp features. 
The way to improve the NN consists in learning harder, that is, we increased the repetitions of the SGD to 100, Fig.~\ref{fig:Fit_function50}b,
and 1000, Fig.~\ref{fig:Fit_function50}c, which finally leads to a good representation of the functions.

A different strategy consist in using different gradient search strategies,
i.e.\ a different optimizer to minimize the cost function Eq.~\eqref{eq:cost}.
In Fig.~\ref{fig:Fit_function50}d we show the results where we used only 500 repetitions, however we switched between
a SGD and an ADAM optimizer, which performs much better, that just an SGD alone. We would like to remark that a priory it is not
clear which optimizer is the best, and the optimizer performance seems to be rather problem dependent, see \cite{Nielsen:DP2015}.

 \begin{figure}
  \begin{center}
  \includegraphics[width=0.6\textwidth,valign=t]{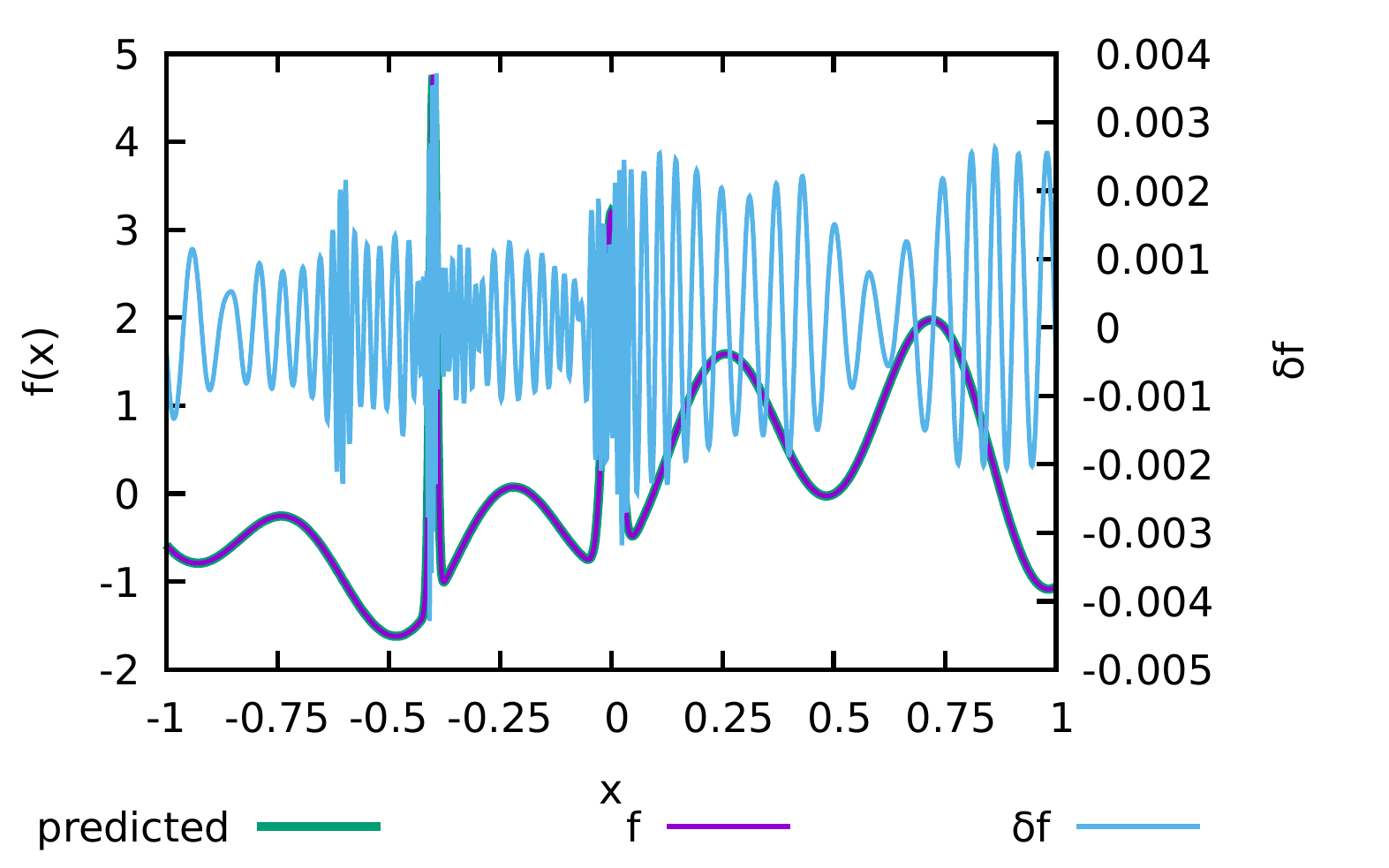}
  \end{center}
\caption{\label{fig:Fit_function250} Fits for the function $f(x)$ Eq.~\eqref{eq:f} using a $\tanh$ activation function obtained
 using {\tt tensorflow/keras} for a $ 1 \times 250 \times 50 \times 50 \times 1$ network,
 250 repetitions, (SGD \& ADAM).
}\end{figure}

Finally in Fig.~\ref{fig:Fit_function250} we present results obtained from a deeper network consisting of
$1 \times 250 \times 50 \times 50 \times 1$ neurons. On the right axis we show the actual error of the fit
which is below $5\,10^{-3}$ over the complete range. In result we obtained  results a rather good approximation
to the function at the expense of using more than 15.000 fit parameter $\{b_{\ell,n},  w_{\ell,n,j}\}$.

We would like to point out that the approach of using 15.000 fit parameter may appear odd as it renders
an understanding of the network impossible. However, we are using the approach to construct a DFT functional. For the ladder
it is also fair to state that most users of the modern sophisticated  DFT functionals have no understanding
on the details of their construction.

\section{A convolutional network}
\label{sec:CNN}
We also tested the setup of a convolutional network. There, in addition to full layer, one constructs a small kernel layer
that gets convoluted with the with the output of another layer. For details we refer to \cite{Nielsen:DP2015}.

\begin{figure}[ht]
\begin{center}
  \begin{minipage}[r]{0.6\textwidth}
        \includegraphics[width=0.5\textwidth]{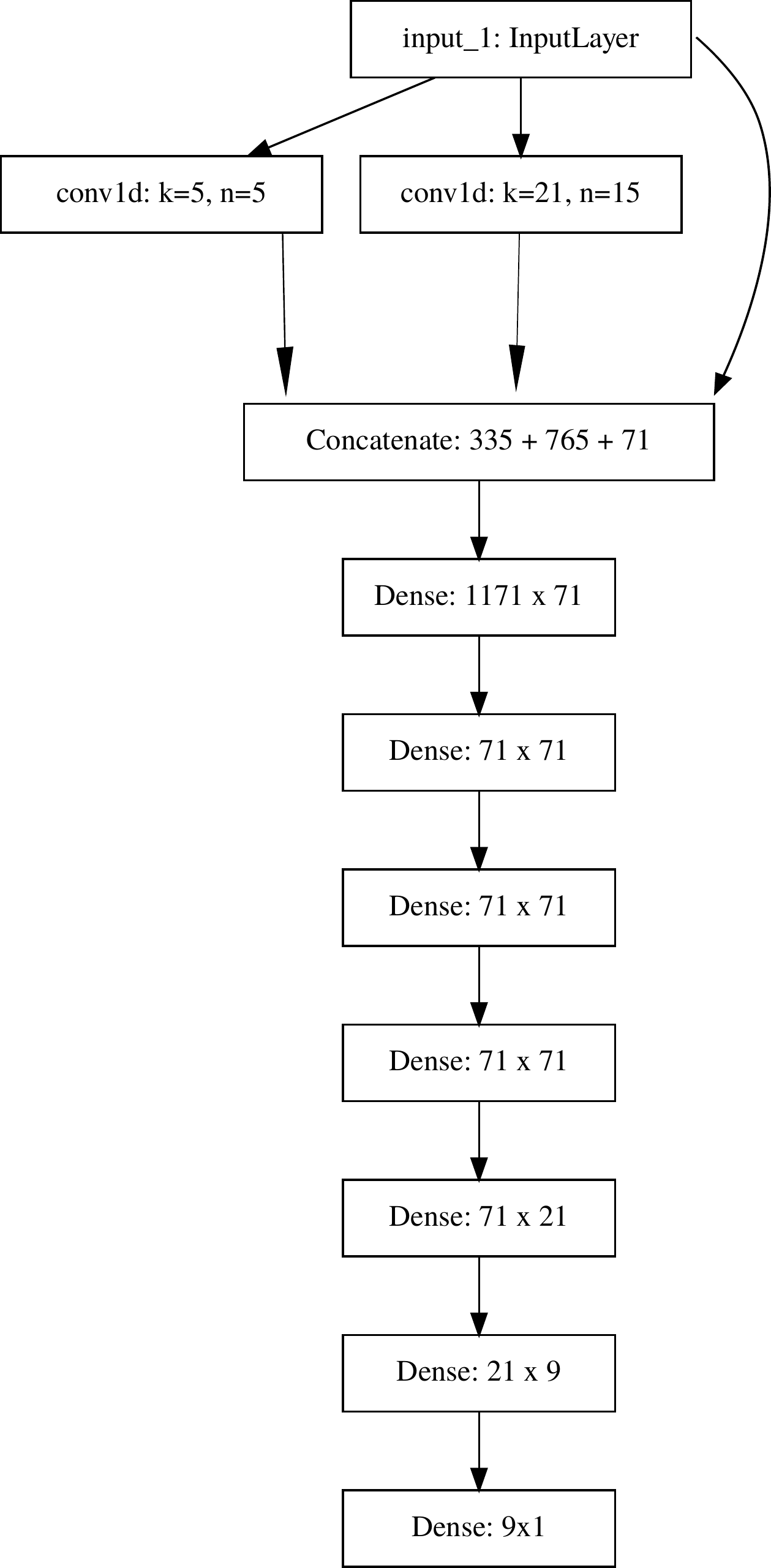}
   \end{minipage}
   \begin{minipage}[l]{0.3\textwidth}
        \caption{Layout for the convolutional network.
        The input layer is connected to two convolutional layer, which are then combined
        with the input layer to serve for the input of seven hidden full layers.
        \label{fig:CNN}}
   \end{minipage}
\end{center}
\end{figure}

Specifically we implemented the a NN as displayed in Fig.~\ref{fig:CNN},
which resulted into 100.628 fit parameter.
However, despite all the effort we could not improve 
on the results obtained from the (smaller) dense network. 

\end{document}